# Polarization Effects on Thermal-Induced Mode Instabilities in High Power Fiber Lasers

Rumao Tao, Pengfei Ma, Xiaolin Wang*, Pu Zhou** and Zejin Liu

*College of Optoelectric Science and Engineering, National University of Defense Technology,
Changsha, Hunan 410073, China*
e-mail: *chinawxllin@ 163.com; **zhoupu203@ 163.com

*Abstract*—We present detailed studies of the effect of polarization on thermal-induced mode instability (MI) in ytterbium-doped fiber amplifiers. Based on a steady-state theoretical model, which takes both electric fields along the two principal axes into consideration, the effect of polarization effects on the gain of Stokes wave was analyzed, which shows that the polarization characteristics of the fiber laser have no impact on the threshold of MI. Experimental validation of the theoretical analysis is presented with experimental results agreeing well with the theoretical results, in which polarization-maintained and non-polarization-maintained fiber lasers with core/inner cladding diameter of 30/250um and core NA of 0.07 were employed. The MI threshold power is measured to be about 367~386W.

*Index Terms*—Fiber laser, polarization, mode instabilities, thermal effects

## I. Introduction

MANY applications, such as coherent lidar system, nonlinear frequency conversion, coherent beam combining architectures, require high power laser sources with diffraction-limited beam quality. Fiber laser systems, which have earned a solid reputation as a highly power scalable laser with high beam quality, are attractive sources for the aforementioned applications [1-3]. Generally, high power fiber laser systems employ large mode area (LMA) fibers to overcome the limitation of nonlinear effects and enable higher power scaling [4]. LMAs results in the onset of thermal-induced mode instabilities (MI), which currently limits the further power scaling of ytterbium doped fiber laser systems with diffraction-limited beam quality and is under intense study [5-19]. Influence of MI on various parameters, i.e. core diameter [6], doped area [8, 10], pump cladding diameter [8, 15], pump/signal wavelength [11, 12, 19], have already been studied in detail. Nevertheless, polarization is another physical character of fiber laser to which little attention has been paid but which should not be ignored. Although the effect of polarization on Stimulated Brillouin Scattering and Stimulated Raman Scattering has been studied in detail [21, 22, 23], no detail theoretical or experimental study on the effect of polarization on MI has been carried out until now. Experimental result has been reported in [12], but technical details are hardly described in the literature.

In this paper, we investigated the effect of polarization on MI theoretically and experimentally. By taking both electric fields along the two principal axes into consideration and the model in [19], we built a steady-state theoretical model to analyze the polarization effects, and the gain of Stokes wave along different principal axes is compared in relation to polarization maintained (PM) and non- PM fiber case. More importantly, a high power master oscillator power amplifier based on PM and non-PM fiber with core/inner cladding diameter of 30/250um has been setup, which enable us to compare experimental results to the theoretical predictions of our theoretical model for the polarization effect case. We find well agreement between theoretical predication and experimental results.

## II. Theoretical Study

The optical field of the signal propagating in the birefringent fiber is expressed in the conventional LP mode representations

$$\mathbf{E}(r,\phi,z,t) = \left[\sum_{m=0}^{\infty}\sum_{n=1}^{\infty} A_{mn}(z,t)\psi_{mn}(r,\phi)e^{j(\beta_{xmn}z-\omega_{mn}t)} + c.c.\right]\vec{e}_x + \left[\sum_{m=0}^{\infty}\sum_{n=1}^{\infty} B_{mn}(z,t)\psi_{mn}(r,\phi)e^{j(\beta_{ymn}z-\omega_{mn}t)} + c.c.\right]\vec{e}_y \quad (1)$$

where *m* and *n* is azimuthal and radial mode numbers respectively. $A_{mn}(z,t)$, $B_{mn}(z,t)$, $\beta_{xmn}$, $\beta_{ymn}$ and $\psi_{mn}(r,\phi)$ are slowly varying mode amplitudes, propagation constants, and normalized mode profiles of $LP_{mn}$ mode. $A_{mn}(z,t)$ and $B_{mn}(z,t)$ are the complex amplitudes along the two principal axes of the birefringent fiber. Assuming the case that the fiber amplifiers are operating below or near the MI threshold, we therefore include only the fundamental mode ($LP_{01}$) and one of the two degenerate $LP_{11}$ modes and the subscripts of 01 and 11 are replaced with 1 and 2 for $LP_{01}$ mode and $LP_{11}$ mode, respectively. Then the signal intensity $I_s$ can be written as

$$I_s(r,\phi,z,t) = 2n_0\varepsilon_0 c\mathbf{E}(r,\phi,z,t)\mathbf{E}(r,\phi,z,t)^* \quad (2)$$
$$= I_0 + \tilde{I}$$

with

$$I_0 = [I_{A11}(z,t) + I_{B11}(z,t)]\psi_1(r,\phi)\psi_1(r,\phi),$$
$$+ [I_{A22}(z,t) + I_{B22}(z,t)]\psi_2(r,\phi)\psi_2(r,\phi)$$
$$\tilde{I} = I_{A12}(z,t)\psi_1(r,\phi)\psi_2(r,\phi)e^{j(q_x z - \Omega t)}$$
$$+ I_{A21}(z,t)\psi_1(r,\phi)\psi_2(r,\phi)e^{-j(q_x z - \Omega t)}, \quad (3)$$
$$+ I_{B12}(z,t)\psi_1(r,\phi)\psi_2(r,\phi)e^{j(q_y z - \Omega t)}$$
$$+ I_{B21}(z,t)\psi_1(r,\phi)\psi_2(r,\phi)e^{-j(q_y z - \Omega t)}$$
$$I_{Akl}(z,t) = 4n_0 \varepsilon_0 c A_k(z,t) A_l^*(z,t),$$
$$I_{Bkl}(z,t) = 4n_0 \varepsilon_0 c B_k(z,t) B_l^*(z,t),$$
$$q = \beta_1 - \beta_2, \quad \Omega = \omega_1 - \omega_2$$

The temperature distribution is governed by heat transportation equation, which is given by

$$\nabla^2 T(r,\phi,z,t) + \frac{Q(r,\phi,z,t)}{\kappa} = \frac{1}{\alpha} \frac{\partial T(r,\phi,z,t)}{\partial t} \quad (4)$$

where $\alpha = \kappa/\rho C$, $\rho$ is the density, $C$ is the specific heat capacity, and $\kappa$ is the thermal conductivity. Since the heat in high power fiber amplifiers is mainly generated from the quantum defect and absorption, the volume heat-generation density $Q$ can be approximately expressed as

$$Q(r,\phi,z,t) \cong g(r,\phi,z,t)\left(\frac{v_p - v_s}{v_s}\right) I_s(r,\phi,z,t) \quad (5)$$

and $g(r,\phi,z,t)$ is the gain of the amplifier

$$g(r,\phi,z,t) = \left[(\sigma_s^a + \sigma_s^e) n_u(r,\phi,z,t) - \sigma_s^a\right] N_{Yb}(r,\phi) \quad (6)$$

where $v_{p(s)}$ is the optical frequencies, $\sigma_s^a$ and $\sigma_s^e$ are the signal absorption and emission cross sections, $\sigma_p^a$ and $\sigma_p^e$ are the pump absorption and emission cross sections, $N_{Yb}(r,\phi)$ is the doping profile, $n_u$ is the steady-state population inversion [13].

Assume that the fiber is water cooled, the appropriate boundary condition for the heat equation at the fiber surface is

$$\kappa \frac{\partial T}{\partial r} + h_q T = 0 \quad (7)$$

where $h_q$ is the convection coefficient for the cooling fluid. By adopting the integral-transform technique to separate variables in the cylindrical system [20], (4), combined with (5) and (6), can be solved as

$$T(r,\phi,z,t) = \frac{1}{\pi} \frac{\alpha n_2}{\eta} \sum_v \sum_{m=1}^{\infty} \frac{R_v(\delta_m, r)}{N(\delta_m)} \int_{t'=0}^{t} e^{-\alpha \delta_m^2 (t-t')} dt'$$
$$\times \begin{Bmatrix} B_{11}(\phi,z)[I_{A11}(z,t') + I_{B11}(z,t')] + B_{22}(\phi,z)[I_{A22}(z,t') + I_{B22}(z,t')] \\ + B_{12}(\phi,z) I_{A12}(z,t') e^{j(q_x z - \Omega t')} + B_{12}(\phi,z) I_{A12}^*(z,t') e^{-j(q_x z - \Omega t')} \\ + B_{12}(\phi,z) I_{B12}(z,t') e^{j(q_y z - \Omega t')} + B_{12}(\phi,z) I_{B12}^*(z,t') e^{-j(q_y z - \Omega t')} \end{Bmatrix} \quad (8)$$

with

$$B_{kl}(\phi,z) = \quad (9a)$$
$$\begin{cases} \int_0^{2\pi} d\phi' \int_0^R g_0 R_v(\delta_m, r') \cos v(\phi - \phi') \frac{\psi_k(r',\phi')\psi_k(r',\phi')}{1 + I_0/I_{saturation}} dr', & k = l \\ \int_0^{2\pi} d\phi' \int_0^R g_0 R_v(\delta_m, r') \cos v(\phi - \phi') \frac{\psi_k(r',\phi')\psi_l(r',\phi')}{(1 + I_0/I_{saturation})^2} dr', & k \neq l \end{cases}$$

$$N(\delta_m) = \int_0^R r R_v^2(\delta_m, r) dr, \quad n_2 = \eta(v_p - v_s)/\kappa v_s \quad (9b)$$

where $v = 0, 1, 2, 3\ldots$ and replace $\pi$ by $2\pi$ for $v = 0$, $\eta$ is the thermal-optic coefficient, $R$ is the radius of the inner cladding, $g_0$ is the small signal gain and $I_{saturation}$ is the saturation intensity. $R_v(\delta_m, r)$ is given by $R_v(\delta_m, r) = J_v(\delta_m r)$ and $\delta_m$ is the positive roots of $\delta_m J_v'(\delta_m R) + J_v(\delta_m R) h_q/\kappa = 0$. Considering effective refractive index of gain from amplifier, the total refractive index, which attributes to gain ($n_g \leq n_0$) and nonlinearity ($n_{NL} \leq n_0$), can be expressed as

$$n^2 = (n_0 + n_g + n_{NL})^2 \cong n_0^2 - j\frac{g(r,\phi,z,t) n_0}{k_0} + 2n_0 n_{NL} \quad (10)$$

where $n_{NL}$ is given by

$$n_{NL}(r,\phi,z,t) = \eta T(r,\phi,z,t)$$
$$= h_{11}(r,\phi,z,t) + h_{22}(r,\phi,z,t) \quad (11)$$
$$+ h_{A12}(r,\phi,z,t) e^{jq_x z} + h_{A21}(r,\phi,z,t) e^{-jq_x z}$$
$$+ h_{B12}(r,\phi,z,t) e^{jq_y z} + h_{B21}(r,\phi,z,t) e^{-jq_y z}$$

with

$$h_{Xkl}(r,\phi,z,t) =$$

$$\begin{cases} \dfrac{\alpha n_2}{\pi} \sum_v \sum_{m=1}^{\infty} \dfrac{R_v(\delta_m, r)}{N(\delta_m)} \int_0^t B_{kk}(\phi,z) I_{Xkk}(z,t') e^{-\alpha \delta_m^2 (t-t')} dt', k = l \\ \dfrac{\alpha n_2}{\pi} \sum_v \sum_{m=1}^{\infty} \dfrac{R_v(\delta_m, r)}{N(\delta_m)} \int_0^t B_{kl}(\phi,z) I_{Xkl}(z,t') e^{-\alpha \delta_m^2 (t-t') - j\Omega t'} dt', k \neq l \end{cases} \quad (12)$$

Inserting (1) and (10) into the wave equation, after very tedious but straightforward derivations, we can obtained the steady-state coupled-mode equations

$$\frac{\partial |A_2|^2}{\partial z} = \left[ \iint g(r,\phi,z)\psi_2\psi_2 r dr d\phi + \left( |A_1|^2 \chi_1(\Omega) + A_1 B_1^* \chi_{21}(\Omega) \right) \right] |A_2|^2 \quad (13a)$$

$$\frac{\partial |B_2|^2}{\partial z} = \left[ \iint g(r,\phi,z)\psi_2\psi_2 r dr d\phi + \left( |B_1|^2 \chi_1(\Omega) + A_1^* B_1 \chi_{22}(\Omega) \right) \right] |B_2|^2 \quad (13b)$$

with

$$\chi_1(\Omega) = 2\frac{n_0 \omega_2^2}{c^2 \beta_2} \text{Im}\left( \iint (\bar{h}_{12}) \psi_1 \psi_2 r dr d\phi \right)$$

$$\chi_{21}(\Omega) = 2\frac{n_0 \omega_2^2}{c^2 \beta_2} \text{Im}\left( e^{i\Delta kz} \iint (\bar{h}_{12}) \psi_1 \psi_2 r dr d\phi \right)$$

$$\chi_{22}(\Omega) = 2\frac{n_0 \omega_2^2}{c^2 \beta_2} \text{Im}\left( e^{-i\Delta kz} \iint (\bar{h}_{12}) \psi_1 \psi_2 r dr d\phi \right) \quad (14)$$

$$\bar{h}_{kl}(r,\phi,z) = \frac{\alpha n_2}{\pi} \sum_v \sum_{m=1}^{\infty} \frac{R_v(\delta_m, r)}{N(\delta_m)} \frac{B_{kl}(\phi,z)}{\alpha \delta_m^2 - j\Omega}$$

$$\Delta k = q_x - q_y \approx 2\pi \delta n \Omega / c$$

where $\delta n$ is the fiber birefringence.

Generally, the frequency shift in MI is on the order of kHz [5-7] and the gain fiber length of the high power fiber laser is about a few meters, which results that the polarization length $l_p = c/\delta n\Omega$ is far larger than the active fiber length $L$ in fiber laser and $\chi(\Omega) = \chi_1(\Omega) = \chi_{21}(\Omega) = \chi_{22}(\Omega)$. If the fiber is excited along one of the principal axes, for example, $I_{A1}^2 = P$ and $I_{B1}^2 = 0$, the gain is $P\chi(\Omega)$ for a Stokes wave polarized in the $x$ direction and 0 for the $y$ direction. Similar to the assumption in [16], we chose equal excitation by a linearly polarized seeding to model non-polarization maintained (PM) case. With $I_{A1}^2 = P/2$ and $I_{B1}^2 = P/2$, the gain is also $P\chi(\Omega)$ for Stokes waves polarized in both direction, which means that there is no difference in threshold for the two excitation conditions. It shows that the polarization characteristics of the fiber laser have no impact on the threshold of MI.

## III. Experimental Study

To verify the theoretical study, both PM and non-PM amplifiers were tested with different seed lasers. First, a 1064nm PM single frequency laser, which was broaden to the linewidth of 2GHz through phase modulation (Pm) and boosted to ~10W by two stages preamplifiers, was employed to seed the PM amplifier as shown in Fig. 1. The main PM amplifier employed a 3m long of PM 30/250 LMA ytterbium-doped fiber (YDF). The core NA of the fiber is about 0.07. The reason to use fiber with aforementioned parameters is to rule out the influence of bend-induced mode specific loss, which can improve the threshold [24, 25]. Six multimode fiber pigtailed 975 nm laser diodes (LD) are used to pump the gain fiber through a (6+1)×1 PM signal/pump combiner. A 1.5 m long double clad Ge-doped fiber (GDF) of the same core/inner cladding diameter and NA as the LMA YDF is spliced to the end of the LMA YDF for power delivery. The spliced region is covered with high-index gel, which acts as cladding mode striper (CMS) to strip the residual pump laser and cladding mode. The output end of the delivery fiber is angle cleaved at 8°. The output laser was collimated by a lens and sampled by a high reflection mirror (HR): the main power part was collected by a power meter while the low power part was imaged on a CCD camera. A Dichroic Mirror (DM) was inserted in the laser path to dump residual pump power. MI was monitored by imaging the laser beam profile with a CCD camera and detecting the time fluctuation of scattering power with photo-detector (PD) [26].

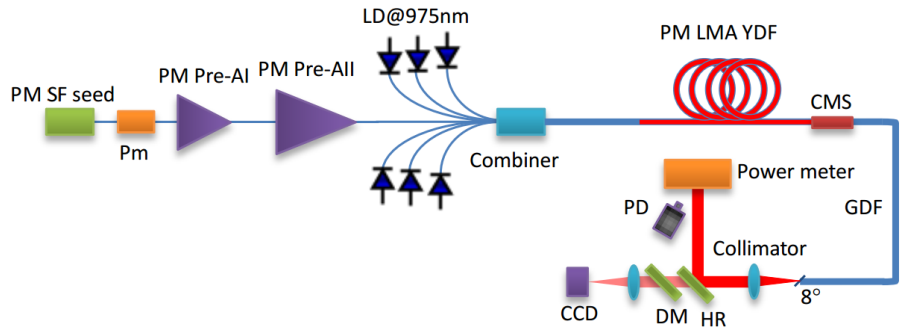

Fig. 1 Experimental setup of the PM amplifier

The threshold is measured to be about 450 Watts, which decreased to 386 Watts after few tests for an estimated one hours of operation including multiple power cycles, and stabilized around 386W. Typical results are presented in Fig. 2. Below the threshold, stable beam profile as shown in Fig. 2(a) was achieved with PER about 96%, which correspond to stable time traces with no fluctuation component (as shown in Fig. 2(c) and (d)). Above the threshold, the beam profiles became instable with fluctuated time traces, and PER also reduced. Frequency component at near 1.7 kHz shown up after the onset of MI, which indicates the time traces in the region exhibit a periodic sawtooth-like oscillation.

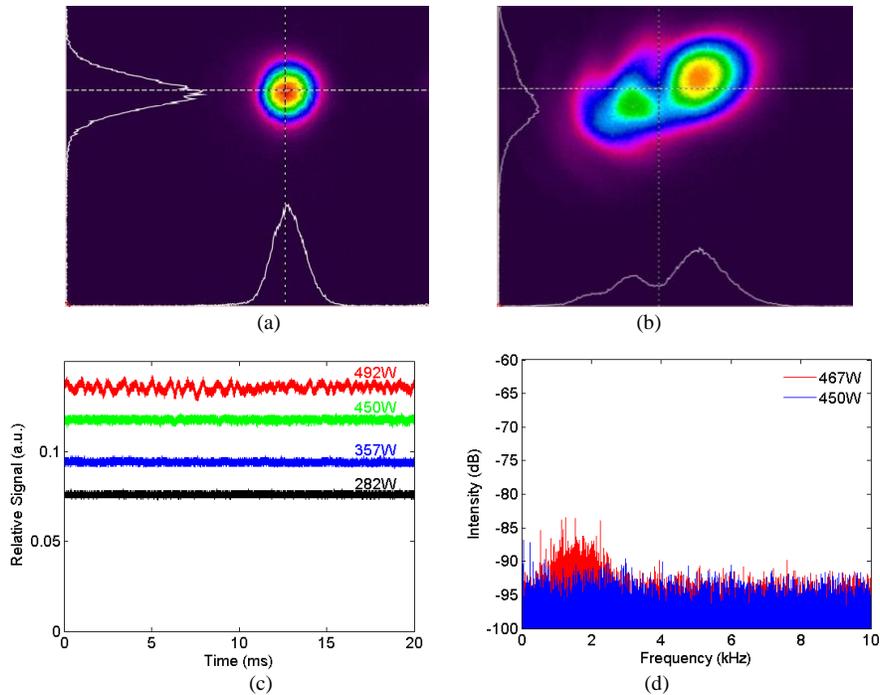

Fig. 2 Characteristics of the PM amplifier. The far-field intensity below (a) and above (b) the MI threshold, (c) are time traces at different output power and (d) are the Fourier spectrogram calculated from time trace at different output power

Then the PM laser with two stage pre-amplifiers was used to seed the non-PM amplifier, which employed a length of non-PM 30/250 LMA gain fiber with the same core NA in the main amplifier. Two cases were studied: case one is with high PER (>95%) while case two is with lower PER (~80%). High PER with non-PM amplifier was achieved by carefully adjusting the polarization direction of the PM laser at the fusion point between the PM fiber of the pre-amplifier and the non-PM fiber of the main amplifier. Similar MI phenomena were observed with nearly the same threshold for both cases, which was lowered from 446W down to 384W after few tests and stabilized around 384W. Frequency component also shown up at near 1.7kHz. The PER as a function of output power is shown in Fig. 3, which shows that, for high PER case, PER reduced dramatically after the onset of MI beyond 384W and became unstable in time. In addition, a non-PM broadband seed @1080nm was also used to seed the same non-PM amplifier, which operating at the output power of 10W with 3dB linewidth of 0.2nm. The threshold is measured to be 367W, which is slightly lower than aforementioned cases. This may due to the wavelength dependence of MI [12]. We can conclude that the experimental results agree with the theoretical prediction and MI is independent of the polarization effects.

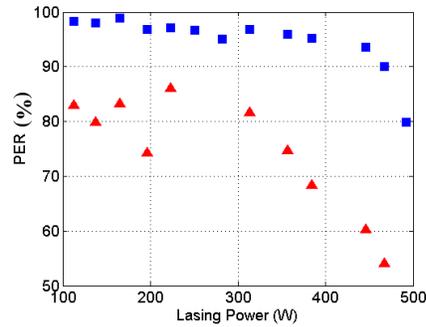

Fig. 3 Polarization evolution of the non-PM amplifier with PM seeding. Red solid triangle is for the case of low PER while blue solid square is for the case of high PER.

## IV. Conclusions

In summary, we have developed a steady-state model to study the effects of polarization on MI. The gain of Stokes wave along different principal axes was analyzed and good agreement between simulations and experimental results was achieved. Both theoretical analysis and experimental study shows that the threshold of MI is the same when linear polarization is maintained as when polarization is scrambled, which means the polarization characteristics of the fiber laser have no impact on MI. Although only one type of fiber was used to examine our theoretical prediction, our model is not limited to special case, which means that our conclusions are universal.